\documentclass[twocolumn,showpacs,preprintnumbers,amsmath,amssymb]{revtex4}

\usepackage{graphicx}
\usepackage{dcolumn}
\usepackage{bm}

\begin{document}

\preprint{APS/123-QED}

\title{The Marginally Stable Circular Orbit of the Fluid Disk around a Black
   Hole}

\author{Lei Qian$^{1}$, Xue-Bing Wu$^{1}$, Li-Xin Li$^{2,1}$}
 \affiliation{$^1$ \ Department of Astronomy, Peking University,
    Beijing, 100871, China\\
      $^2$ \ Kavli Institute for Astronomy \& Astrophysics, Peking
University, Beijing, 100871, China}
%

\date{\today}

\begin{abstract}
The inner boundary of a black hole accretion disk is often set to the
marginally stable circular orbit (or the innermost stable circular
orbit, ISCO) around the black hole. It is important for the theories of
black hole accretion disks and their applications to astrophysical black hole systems.
Traditionally, the marginally stable circular orbit is obtained by considering the
equatorial motion of a test particle around a black hole. However, in reality the
accretion flow around black holes consists of fluid, in which the pressure
often plays an important role. Here we consider the influence of fluid pressure on the
location of marginally stable circular orbit around black holes. It is found that
when the temperature of the fluid is so low that the thermal energy of a particle is
much smaller than its rest energy, the location of marginally stable circular
orbit is almost the same as that in the test particle case. However,
we demonstrate that in some special cases the
marginally stable circular orbit can be different when the fluid pressure is large
and the thermal energy becomes non-negligible comparing with the rest energy.
We present our results for both the cases
of non-spinning and spinning black holes. The influences of
our results on the black hole spin parameter measurement in X-ray binaries and the
energy release efficiency of accretion flows around black holes are discussed.
\end{abstract}

\maketitle

\section{Introduction}           
\label{sect:intro}

In the spacetime around a black hole, the circular motion of a test
particle is not always possible according to general relativity.
There is a smallest radius on which the circular motion of a test
particle is marginally stable \cite{bardeen1972}. This
radius is called marginally stable circular orbit. With
a small perturbation, the circularly moving particle at this radius
will then plunge into the black hole freely. The properties of the
transition from inspiral to plunge depend on the mass ratio $\eta$
of the particle and the black hole \cite{ori2000}. The
transition would be less gradual (or more "abrupt") with smaller
$\eta$ (See also \cite{oshaughnessy2003, sundararajan2008}). This
property is more relevant in the context of the inspiral in a binary
black hole system, when the two black holes are just about to merge.
When the self-force of the inspiralling particle is considered, the
location of the marginally stable orbit is also modified \cite{barack2009}.

The marginally stable circular orbit is also important in the
estimate of energy release of the accretion to a black hole. By
calculating the binding energy of the circular motion on this radius
\cite{salpeter1964}, one can estimate how much of the total
energy can be released during the accretion process. For the
accretion flow around a black hole, the energy release efficiency
rangs from 5.6\% (for non-spinning or Schwarzschild black holes) to
42\% (for extreme Kerr black holes).  The
location of the marginally stable circular orbit and the
energy release efficiency are closely related, both depending on the
spin parameter of the black hole \cite{novikov1973}.

In some applications of accretion disk model in astrophysical
observations, the location of marginally stable circular orbit is
also crucial. Since the marginally stable circular orbit depends on the spin
of the central black hole, it is possible to measure the spin of a
black hole if the corresponding marginally stable orbit can be
measured. By fitting the observed soft state spectra of black hole
X-ray binaries, which are assumed to come from a thin disk with its
inner boundary at the marginally stable circular orbit, one can
derive the location of this orbit and estimate the spin parameters of
the black holes in X-ray binaries \cite{zhang1997}.

In the theory of accretion flow, the marginally stable circular
orbit is also important. Due to its transitional nature from the
inspiralling region to the plunging region, it is often believed
that an accretion disk is torque free at this radius. Although still
being debated when considering magnetic fields \cite{krolik2002,shafee2008}, 
the torque free condition on the marginally stable circular
orbit serves as an important inner boundary condition in many accretion disk models.

The self-gravity of the particle is not relevant in the context of
accretion disks, but the material in a real accretion disk is fluid
rather than test particles. When considering fluid, the particles
within it are interacting with each other, and the pressure plays an
important role in the dynamics. In the study of accretion
disk theory, the marginally stable circular orbit for a test particle
is used as the inner boundary of an accretion disk in many models.
More precisely, it has been shown by analytical theory of fluid tori around
black holes that the inner boundary of an accretion disk
lies between the marginally stable circular orbit and the marginally
bound orbit, while both expressions are for test particles
\cite{fishbone1976,kozlowski1978,abramowicz1978}.
In this work, however, we focus on a different problem, that is,
we try to investigate the marginally stable circular orbit itself,
with the influence from the pressure in the fluid.
In section 2, we briefly mention the marginally stable circular
orbit in the case of a test particle. Then we consider a thin disk consisting of
perfect fluid around a black hole and discuss the marginally stable
circular orbit in this case in section 3. In Section 4 we present a
brief discussion on our results. In this paper, we set $c=G=M=1$,
with $c$, $G$, and $M$ the speed of light, gravitational constant,
and the mass of the black hole, respectively. We use ($+\ -\ -\ -$)
signature all through.

\section{Marginally Stable Circular Orbit of A Test Particle}
\label{sect:particle}

The metric of the spacetime outside a Kerr (spinning, non-charged)
black hole is \cite{bardeen1972}
\begin{equation}
ds^2=\frac{\rho^2\Delta}{A}dt^2-\frac{A\sin^2\theta}{\rho^2}(d\phi-\omega
dt)^2-\frac{\rho^2}{\Delta}dr^2-\rho^2 d\theta^2,
\end{equation}
where
$$
\rho^2=r^2+a^2\cos^2\theta,
$$
$$
A=(r^2+a^2)^2-\Delta a^2\sin^2\theta
$$
\begin{equation}
\Delta=r^2-2r+a^2,\ \ \omega=\frac{2ar}{A},
\end{equation}
and $a$ is the spin parameter of the black hole. The covariant form
of the metric can be written as
$$
g^{tt}=-\frac{1}{\rho^2}\left[a^2\sin^2\theta-\frac{(r^2+a^2)^2}{\Delta}\right],\
\ g^{rr}=-\frac{\Delta}{\rho^2},
$$
$$
g^{\theta\theta}=-\frac{1}{\rho^2},\ \
g^{\phi\phi}=-\frac{1}{\rho^2}\left(\frac{1}{\sin^2\theta}-\frac{a^2}{\Delta}\right),
$$
\begin{equation}
g^{t\phi}=-\frac{a}{\rho^2}\left(1-\frac{r^2+a^2}{\Delta}\right).
\end{equation}

For a steady and axis-symmetric spacetime (e.g. Schwarzschild, Kerr),
there are two apparent constants of motion of a free particle with
unit mass
\begin{equation}
u_{t}=E,\ \ u_{\phi}=-L \label{constant_motion}
\end{equation}
Since the metric is block diagonal, the normalization
condition of the 4-velocity can be written as
\begin{equation}
g^{tt} u_t u_t+2g^{t\phi}u_t
u_{\phi}+g^{\phi\phi}u_{\phi}u_{\phi}+g_{rr}u^r
u^r+g^{\theta\theta}u^{\theta}u^{\theta}=1
\end{equation}
Consider the circular orbital motion on the equatorial plane
($\theta=\pi/2$, $u^{\theta}\equiv d\theta/d\tau=0$), the equation
of motion can be written as
\begin{equation}
r^4\left(\frac{dr}{d\tau}\right)^2=U_r,
\end{equation}
where
\begin{eqnarray*}
U_r&=&(r^4+a^2 r^2+2a^2 r)E^2+4ar EL-(r^2-2r)L^2\\
&&-(r^4-2r^3+a^2 r^2).
\end{eqnarray*}
\begin{equation}
\label{eff_potential}
\end{equation}

In order to maintain a circular orbit, $dr/d\tau$ and $d^2r/d\tau^2$
should both vanish, which imply
\begin{equation}
U_r=0,\ \ U_r'=0,
\end{equation}
where the prime denotes the derivative to $r$. The two constants of
motion can be derived from the above two equations as
\begin{equation}
E=\frac{r^{3/2}-2r^{1/2}+ a}{r^{3/4}(r^{3/2}-3r^{1/2}+ 2a)^{1/2}},
\label{kerr_energy}
\end{equation}
\begin{equation}
L=-\frac{r^2- 2ar^{1/2}+a^2}{r^{3/4}(r^{3/2}-3r^{1/2}+ 2a)^{1/2}}.
\label{kerr_angular_momentum}
\end{equation}
For prograde motion, $a>0$; for retrograde motion, $a<0$. Not all
the circular orbits are stable. The stable ones should also
fulfill $U_r''\le 0$, and the marginal stable circular orbit
corresponds to $U_r''=0$. Using the expression for $E$ and $L$, we
can derive the marginal stable orbit
\begin{equation}
r_{ms}=3+Z_2\mp [(3-Z_1)(3+Z_1+2Z_2)]^{1/2}, \label{kerr_radius}
\end{equation}
where
$$
Z_1\equiv 1+\left(1-a^2\right)^{1/3}[(1+a)^{1/3}+(1-a)^{1/3}],
$$
$$
Z_2\equiv (3a^2+Z_1^2)^{1/2}.
$$
The upper sign and the lower sign are for prograde and retrograde
motion of the particle, respectively. Note that for a Schwarzschild
black hole, $a=0$, $r_{ms}=6$. The expression above has been widely
adopted in various astrophysical studies of black hole systems.

\section{Marginally Stable Circular Orbit of A Perfect Fluid Disk}
\label{sect:fluid}

For perfect fluid, the energy-momentum tensor can be written as
\begin{equation}
T^{\mu}_{\nu}=(p+\varepsilon)u^{\mu}u_{\nu}+p\delta^{\mu}_{\nu},
\end{equation}
where $u^{\mu}$, $p$, and $\varepsilon$ are the 4-velocity,
pressure, and the energy density respectively. It can be proved that
there are also two constants of motion \cite{abramowicz1978}:
\begin{equation}
\frac{p+\varepsilon}{n}u_t=\mathcal{E},
\end{equation}
and
\begin{equation}
\frac{p+\varepsilon}{n}u_{\phi}=-\mathcal{L},
\end{equation}
where $n$ is the number density. Note that these two constants of
motion represent no longer the energy and angular momentum. Use the
normalization of the 4-velocity, it is easy to show that
\begin{equation}
r^4\left(\frac{dr}{d\tau}\right)^2=U_r,
\end{equation}
where the effective potential
\begin{eqnarray*}
U_r&=&\left(\frac{n}{p+\varepsilon}\right)^2[(r^4+a^2 r^2+2a^2
r)\mathcal{E}^2+4ar
\mathcal{EL}\\
&&-(r^2-2r)\mathcal{L}^2]-(r^4-2r^3+a^2 r^2).
\end{eqnarray*}
\begin{equation}
\label{eff_potential2}
\end{equation}

\subsection{The Schwarzschild Black Hole Case}

We first consider the case of a thin fluid disk around a non-rotating black
hole (with spin parameter $a=0$). The effective potential becomes
\begin{equation}
U_r=\left(\frac{n}{p+\varepsilon}\right)^2[r^4\mathcal{E}^2-(r^2-2r)\mathcal{L}^2]-(r^4-2r^3).
\label{eff_potential2_S}
\end{equation}
In this case, the energy density $\varepsilon$ consists of two
parts: One is the rest energy (proportional to the number density
$n$) and the other one is the thermal energy (assumed to
be proportional to the pressure $p$). Therefore we have
\begin{equation}
\varepsilon=m_{_0} n+\frac{1}{\gamma-1}p,
\end{equation}
where $m_0$ and $\gamma$ are the rest mass of the particle and the
ratio of specific heat, both of which are constants. If the
dependence of both number density and pressure on the radius are of power
law forms, we can parameterize the function $n/(p+\varepsilon)$ as
\begin{equation}
\frac{n}{p+\varepsilon}=\frac{B}{1+Cr^{-b}},
\end{equation}
where we have assumed $p/\rho\propto T\propto r^{-b}$, where
$\rho\equiv m_0 n$ is the density of the fluid. For the standard
thin disk model \cite{shakura1973,novikov1973},
$b$ is $3/8$, $9/10$, $3/4$, in the inner, middle and outer
regions, respectively. For the self-similar solution of an advection
dominated accretion flow (ADAF, \cite{narayan1994,narayan1995}), $b$
equals to $1$. However, one should note that these dependence on
radius $r$ is only for the region where $r\gg r_{ms}$. When $r\sim
r_{ms}$, $b$ can be negative if the torque free condition at $r\sim
r_{ms}$ is applied, which may be more relevant in the case of real
accretion flows around black holes. The constant $C$ is always
positive.

Since the constant $B$ can be absorbed into $\mathcal{E}$ and
$\mathcal{L}$, the effective potential can be rewritten as
\begin{equation}
U_r=\left(\frac{1}{1+Cr^{-b}}\right)^2[r^4\mathcal{E}^2-(r^2-2r)\mathcal{L}^2]-(r^4-2r^3).
\label{eff_potential2_SP}
\end{equation}

In a normal fluid, the temperature is usually low and the thermal
energy is much smaller than the rest energy, that is
\begin{equation}
\frac{1}{\gamma-1}p\ll m_{_0} n.
\end{equation}
In another word, the constant $C$ is very small ($C\ll 1$) in this case. Then there is not much
influence from the pressure. The effective potential is
approximately the same as that in the test particle case, which
means the location of the marginally stable circular orbit is almost the same
even in the fluid case.

However, in some hot accretion flows around black holes, e.g. ADAF \cite{narayan1994,narayan1995}, 
the thermal energy may be relevant. We can estimate
the ratio of thermal energy to rest energy in this case as
\begin{equation}
\theta_i\equiv \frac{kT_i}{m_i c^2}\approx 0.1,
\end{equation}
\begin{equation}
\theta_e\equiv \frac{kT_e}{m_e c^2}\approx 0.1,
\end{equation}
where $k$, $T_i$, $m_i$, $T_e$, $m_e$, $c$ are the Boltzmann
constant, ion temperature, ion mass, electron temperature, and speed of
light, respectively. In this case, $C$ is not very large ($C<1$) but also
non-negligible. We
can expand the marginal stable circular orbit as
\begin{equation}
r_{ms}=r_{ms,0}+C\Delta r_{ms}.
\end{equation}

In order to calculate the correction term $\Delta r_{ms}$, we have to know
the first and second derivatives of the effective potential, Eq. (\ref{eff_potential2_SP}).
They are
\begin{eqnarray*}
U_r'&=&\frac{2bCr^{-(b+1)}}{(1+Cr^{-b})^3}[r^4\mathcal{E}^2-(r^2-2r)\mathcal{L}^2]\\
&&+\frac{1}{(1+Cr^{-b})^2}[4r^3\mathcal{E}^2-(2r-2)\mathcal{L}^2]-(4r^3-6r^2)
\end{eqnarray*}
\begin{equation}
\end{equation}
and
\begin{eqnarray*}
U_r''&=&\frac{2b(b+1)Cr^{-(b+2)}}{(1+Cr^{-b})^3}[r^4\mathcal{E}^2-(r^2-2r)\mathcal{L}^2]\\
     & &+\frac{6b^2C^2r^{-2(b+1)}}{(1+Cr^{-b})^4}[r^4\mathcal{E}^2-(r^2-2r)\mathcal{L}^2]\\
     & &+\frac{4bCr^{-(b+1)}}{(1+Cr^{-b})^3}[4r^3\mathcal{E}^2-(2r-2)\mathcal{L}^2]\\
     & &+\frac{1}{(1+Cr^{-b})^2}[12r^2\mathcal{E}^2-2\mathcal{L}^2]-(12r^2-12r).
\end{eqnarray*}
\begin{equation}
\end{equation}
Setting Eq. (\ref{eff_potential2_SP}) and above two
equations to $0$ ($U_r=U_r'=U_r''=0$), we get an equation after
eliminating $\mathcal{E}$ and $\mathcal{L}$,
$$
(r-3)[-\frac{2b(b+1)Cr^{-b}}{1+Cr^{-b}}(r-2)-\frac{2b^2C^2r^{-2b}}{(1+Cr^{-b})^2}(r-2)
$$
$$
+\frac{4bCr^{-b}}{1+Cr^{-b}}(4r-6)]
+(10r-24)\left[1-\frac{bCr^{-b}}{1+Cr^{-b}}(r-2)\right]
$$
\begin{equation}
-12(r-3)=0.
\label{Schwarzschild_expand}
\end{equation}
Keeping the terms of the order $O(C)$  we can get
\begin{eqnarray*}
\Delta r_{ms}&=& b r_{ms,0}^{-b}\{(r_{ms,0}-3)[(7-b)r_{ms,0}+(2b-10)]\\
&&-(r_{ms,0}-2)(5r_{ms,0}-12))\}
\end{eqnarray*}
\begin{equation}
\end{equation}

For a Schwarzschild spacetime, $r_{ms,0}=6$, so we get
\begin{equation}
r_{ms}=r_{ms,0}+C\Delta r_{ms}=6+6^{-b}Cb(24-12b).
\end{equation}

   \begin{figure}[!htbp]
   \centering
   \includegraphics[width=7cm, angle=0]{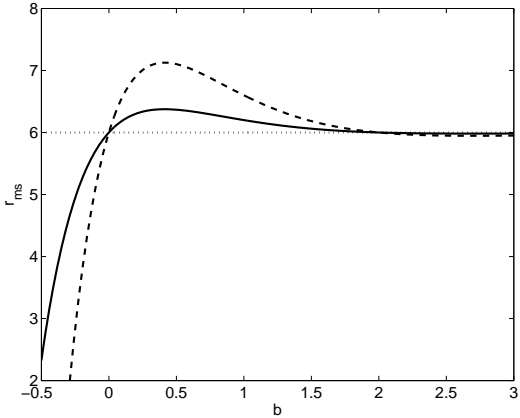}
   \caption{The dependence of the marginally stable circular orbit $r_{ms}$ in Schwarzschild spacetime on
            the power-law index $b$ in the case of $C<1$. The solid line and dashed line are
            for $C=0.1$ and $C=0.3$, respectively. The dotted horizontal line represents $r_{ms}=6$ in the test particle case.}
   \label{figure_rmsb}
   \end{figure}

In the equation above, $\Delta r_{ms}(b)$ has its maximum $3.75$ at $b=0.413$ and local
minimum $-0.18$ at $b=2.7$, respectively. So the maximum of the
marginally stable circular orbit is
\begin{equation}
r_{ms,max}=6+3.75C,
\end{equation}
corresponding to $b=0.413$ and the local minimum is
\begin{equation}
r_{ms,min}=6-0.18C,
\end{equation}
corresponding to $b=2.7$. When $b\to \infty$, $r_{ms}\to 6$, and
when $b$ is negative, $r_{ms}<6$ and it decreases monotonically with
the decreasing of $b$ (see Fig. \ref{figure_rmsb}). Therefore, in this
case when the thermal energy is smaller than the rest energy, the
influence of the pressure on the marginally stable circular orbit is
non-negligible, especially when $b<1.5$.

\subsection{The Kerr Black Hole Case}
In Kerr spacetime, the effective potential is expressed by Eq.
(\ref{eff_potential2}). When $C$ is small, we can do the same
expansion as in Schwarzschild spacetime.

In Kerr spacetime, the effective potential, and its first and second
order derivatives are
\begin{eqnarray*}
U_r&=&\left(\frac{1}{1+Cr^{-b}}\right)^2[(r^4+a^2
r^2+2a^2r)\mathcal{E}^2+4ar\mathcal{EL}\\
&&-(r^2-2r)\mathcal{L}^2]-(r^4-2r^3+a^2r^2)
\label{eff_potential2_KP}
\end{eqnarray*}
\begin{equation}
\end{equation}
$$
U_r'=\frac{2bCr^{-(b+1)}}{(1+Cr^{-b})^3}[(r^4+a^2
r^2+2a^2r)\mathcal{E}^2+4ar\mathcal{EL}-(r^2-2r)\mathcal{L}^2]+
$$
$$
\frac{1}{(1+Cr^{-b})^2}[(4r^3+2a^2r+2a^2)\mathcal{E}^2+4a\mathcal{EL}-(2r-2)\mathcal{L}^2]
$$
\begin{equation}
-(4r^3-6r^2+2a^2r)
\end{equation}
and
\begin{eqnarray*}
U_r''&=&-\frac{2b(b+1)Cr^{-(b+2)}}{(1+Cr^{-b})^3}[(r^4+a^2
r^2+2a^2r)\mathcal{E}^2+4ar\mathcal{EL}\\
     & &-(r^2-2r)\mathcal{L}^2]\\
     & &+\frac{6b^2C^2r^{-2(b+1)}}{(1+Cr^{-b})^4}[(r^4+a^2
     r^2+2a^2r)\mathcal{E}^2+4ar\mathcal{EL}\\
     & &-(r^2-2r)\mathcal{L}^2]+\frac{4bCr^{-(b+1)}}{(1+Cr^{-b})^3}[(4r^3+2a^2r+2a^2)\mathcal{E}^2\\
     & &+4a\mathcal{EL}-(2r-2)\mathcal{L}^2]+\frac{1}{(1+Cr^{-b})^2}[(12r^2+2a^2)\mathcal{E}^2\\
     & &-2\mathcal{L}^2]-(12r^2-12r+2a^2)
\end{eqnarray*}
\begin{equation}
\end{equation}
respectively. After some long but straightforward deductions, we
can get
\begin{widetext}
\begin{equation}
\Delta
r_{ms}=\frac{2b(b+3)r_0^{-b}(r_0^2-2r_0+a^2)-4br^{-b}(4r_0^2-6r_0+2a^2)-6r_0^2
\mathcal{E}^2_1}{4+12r_0\mathcal{E}_0^2+12r_0^2\mathcal{E}_0\mathcal{E}_0'-12r_0},
\end{equation}
\end{widetext}
where $r_0$ and $\mathcal{E}_0$ are expressed as
\begin{equation}
r_{0}=3+Z_2\mp [(3-Z_1)(3+Z_1+2Z_2)]^{1/2}, \label{kerr_radius0}
\end{equation}
and
\begin{equation}
\mathcal{E}_0=\frac{r_0^{3/2}-2r_0^{1/2}+
a}{r_0^{3/4}(r_0^{3/2}-3r_0^{1/2}+ 2a)^{1/2}}, \label{kerr_energy0}
\end{equation}
respectively, where
$$
Z_1\equiv 1+\left(1-a^2\right)^{1/3}[(1+a)^{1/3}+(1-a)^{1/3}],
$$
$$
Z_2\equiv (3a^2+Z_1^2)^{1/2}.
$$
$\mathcal{E}_0'$ is the derivative of $\mathcal{E}_0$ with $r_0$.
$\mathcal{E}^2_1$ is the correction term of $\mathcal{E}_0^2$ with
order $O(C)$.

\begin{figure}[!htbp]
   \centering
   \includegraphics[width=7cm, angle=0]{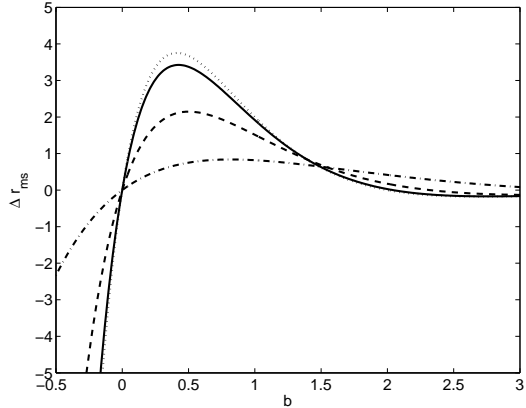}
   \caption{The dependence of the correction term $\Delta r_{ms}$ on the power-law index $b$ and  spin parameter $a$.
            Dotted line: $a=0$; solid line: $a=0.1$; dashed line: $a=0.5$; dash-dotted line:
            $a=0.9$.}
   \label{figure_rmsb3}
   \end{figure}

\begin{figure}[!htbp]
   \centering
   \includegraphics[width=7cm, angle=0]{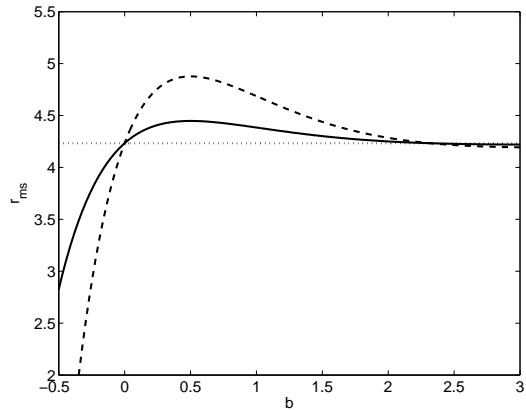}
   \caption{The dependence of the marginally stable circular orbit $r_{ms}$ in Kerr spacetime with $a=0.5$ on
            the power-law index $b$ in the case of $C<1$. The solid line and dashed line are
            for $C=0.1$ and $C=0.3$, respectively. The dotted horizontal line represents $r_{ms}=4.233$ in the test particle case for $a=0.5$.}
   \label{figure_rmsbk}
   \end{figure}

The behavior of the correction term is similar to that of the
Schwarzschild case. This can be seen in Fig. \ref{figure_rmsb3}. As
an example, Fig. \ref{figure_rmsbk} shows the dependence of the
marginally stable circular orbit on $b$ and $C$ in Kerr spacetime
with $a=0.5$.

\section{Discussion}
\label{sect:discussion}

The marginally stable circular orbit can be treated as the
inner boundary of an accretion disk around a black hole in some cases.
This is important because it is crucial to study the structure of accretion
disks. It is also important when calculating the radiation of a standard
accretion disk, where it is usually assumed that there is no
radiation from the region inside the marginally stable circular orbit.

However, the widely accepted marginally stable circular orbit is
based on the test particle assumption. When the temperature in an
accretion disk is not very high (namely the thermal energy is much smaller than
rest energy), the expression of the marginally stable circular orbit for a test particle
is accurate enough. But when the temperature is so high that the thermal energy is
non-negligible comparing to the rest energy, pressure may introduce some
corrections to the location of the marginally stable circular orbit.

If the torque free boundary condition is applied at the inner boundary
of accretion disks, the temperature near the marginally stable circular orbit
would decrease with the decreasing of radius, which is different from the region
far from the inner boundary. Assuming the dependence of temperature on
radius near the marginally stable circular orbit is a power-law
$T\propto r^{-b}$, then $b<0$. As can be seen from Fig.
\ref{figure_rmsb3}, the correction term changes significantly with
$b$ when $b$ is negative. As an example, for the Schwarzschild case,
when $C=0.2$ and $b=-0.2$, the marginally stable circular orbit is
$r_{ms}=5$.

\begin{figure}[!htbp]
   \centering
   \includegraphics[width=7cm, angle=0]{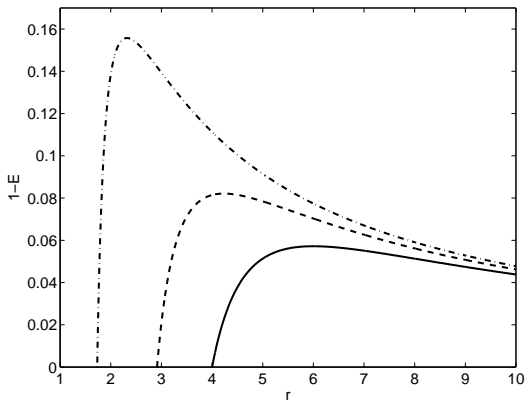}
   \caption{The dependence of the efficiency $1-E$ on radius $r$. The solid line, dashed line anddash-dotted line correspond to the cases of $a=$0, 0.5 and 0.9, respectively.}
   \label{figure_energy_S}
   \end{figure}

The correction to the marginally stable circular orbit not only
influence the inner boundary of accretion disks, but also affect
the energy release of the accretion flow around black holes. In Schwarzschild
spacetime, the binding energy $E$ (see Eq. \ref{kerr_energy})
at a certain radius $r$ is \cite{bardeen1972}
\begin{equation}
E=\frac{r-2}{r^{1/2}(r-3)^{1/2}},
\end{equation}
which equals to $0.9428$ at $r=r_{ms}=6$, so the corresponding efficiency is
$1-0.9428=5.72\%$. If the marginally stable circular orbit is
different, the efficiency can be different, e.g., at $r=5$ the
efficiency is $5.13\%$. In Kerr spacetime, the energy $E$ of a test
particle co-rotating in a circular orbit at $r$ is described by Eq.
(\ref{kerr_energy})
The dependence of the efficiency $1-E$ on radius $r$ is shown in
Fig. \ref{figure_energy_S}. Note that it is not monotonic, when the
radius $r<r_{ms}$, the efficiency is also lower than that at the
marginally stable circular orbit.

However, not only the true location of inner boundary of accretion disks
is different from marginally stable circular orbit as shown
by the analytic models \cite{fishbone1976,kozlowski1978,abramowicz1978},
but as \cite{mineshige2005} also mentioned, in a disk
where the advection is important, it is not proper to use the
marginally stable circular orbit as its inner boundary to calculate
the radiation of the disk. The reason is that in an advection
dominated flow, there is no "equilibrium" of the circular motion,
and the condition based on which the marginally stable circular
orbit is derived is not fulfilled. But the widely used
technique to measure black hole spin by fitting thermal soft state
spectra \cite{zhang1997,shafee2006} is all right
because in a standard thin disk, the marginally stable orbit is
almost the same as for the test particle case, which is used for the
measurement of the black hole spin parameters in X-ray binaries.

One thing should be noted is that our treatment is based on a
perfect fluid disk. However, a real accretion disk consists of
viscous fluid. Another point that may worth further consideration is the
influence of pressure on the "abruptness" of the transition from
inspiral to plunge regions. Because this property will heavily
affect the inner boundary condition of accretion disks, if the
transition is somehow less "abrupt", the fluid disk may continue
further in beyond the marginally stable circular orbit. However,
this is out of the scope of this work. We hope to tackle this
problem in a future work.

\begin{acknowledgments}
This work is supported by the NSFC grant 10525113 and the 973
program 2007CB815405. Q.L. would like to thank all the members of
AGN group in PKU for helpful discussions.
\end{acknowledgments}

\appendix

\newpage 
\bibliography{ms}

\begin{thebibliography}{18}
\expandafter\ifx\csname natexlab\endcsname\relax\def\natexlab#1{#1}\fi
\expandafter\ifx\csname bibnamefont\endcsname\relax
  \def\bibnamefont#1{#1}\fi
\expandafter\ifx\csname bibfnamefont\endcsname\relax
  \def\bibfnamefont#1{#1}\fi
\expandafter\ifx\csname citenamefont\endcsname\relax
  \def\citenamefont#1{#1}\fi
\expandafter\ifx\csname url\endcsname\relax
  \def\url#1{\texttt{#1}}\fi
\expandafter\ifx\csname urlprefix\endcsname\relax\def\urlprefix{URL }\fi
\providecommand{\bibinfo}[2]{#2}
\providecommand{\eprint}[2][]{\url{#2}}

\bibitem[{\citenamefont{{J. M. {Bardeen}, W. H. {Press}, S. A.
  {Teukolsky}}}(1972)}]{bardeen1972}
\bibinfo{author}{\bibnamefont{{J. M. {Bardeen}, W. H. {Press}, S. A.
  {Teukolsky}}}}, \bibinfo{journal}{ApJ} \textbf{\bibinfo{volume}{178}},
  \bibinfo{pages}{347} (\bibinfo{year}{1972}).

\bibitem[{\citenamefont{{A. {Ori} \& K. S. {Thorne}}}(2000)}]{ori2000}
\bibinfo{author}{\bibnamefont{{A. {Ori} \& K. S. {Thorne}}}},
  \bibinfo{journal}{Phys. Rev. D} \textbf{\bibinfo{volume}{62}},
  \bibinfo{pages}{124022} (\bibinfo{year}{2000}).

\bibitem[{\citenamefont{{R. {O'shaughnessy}}}(2003)}]{oshaughnessy2003}
\bibinfo{author}{\bibnamefont{{R. {O'shaughnessy}}}}, \bibinfo{journal}{Phys.
  Rev. D} \textbf{\bibinfo{volume}{67}}, \bibinfo{pages}{04004}
  (\bibinfo{year}{2003}).

\bibitem[{\citenamefont{{P. A. {Sundararajan}}}(2008)}]{sundararajan2008}
\bibinfo{author}{\bibnamefont{{P. A. {Sundararajan}}}}, \bibinfo{journal}{Phys.
  Rev. D} \textbf{\bibinfo{volume}{77}}, \bibinfo{pages}{124050}
  (\bibinfo{year}{2008}).

\bibitem[{\citenamefont{{L. {Barack} \& N. {Sago}}}(2009)}]{barack2009}
\bibinfo{author}{\bibnamefont{{L. {Barack} \& N. {Sago}}}}
  (\bibinfo{year}{2009}), \eprint{{0902.0573}}.

\bibitem[{\citenamefont{{E. E. {Salpeter}}}(1964)}]{salpeter1964}
\bibinfo{author}{\bibnamefont{{E. E. {Salpeter}}}}, \bibinfo{journal}{ApJ}
  \textbf{\bibinfo{volume}{140}}, \bibinfo{pages}{796} (\bibinfo{year}{1964}).

\bibitem[{\citenamefont{{Novikov} and {Thorne}}(1973)}]{novikov1973}
\bibinfo{author}{\bibfnamefont{I.~D.} \bibnamefont{{Novikov}}}
  \bibnamefont{and} \bibinfo{author}{\bibfnamefont{K.~S.}
  \bibnamefont{{Thorne}}}, in \emph{\bibinfo{booktitle}{Black Holes (Les Astres
  Occlus)}} (\bibinfo{year}{1973}), pp. \bibinfo{pages}{343--450}.

\bibitem[{\citenamefont{{S.-N. {Zhang}, W. {Cui}, W.
  {Chen}}}(1997)}]{zhang1997}
\bibinfo{author}{\bibnamefont{{S.-N. {Zhang}, W. {Cui}, W. {Chen}}}},
  \bibinfo{journal}{ApJ} \textbf{\bibinfo{volume}{482}}, \bibinfo{pages}{L155}
  (\bibinfo{year}{1997}).

\bibitem[{\citenamefont{{J. H. {Krolik} \& J. F. {Hawley}}}(2002)}]{krolik2002}
\bibinfo{author}{\bibnamefont{{J. H. {Krolik} \& J. F. {Hawley}}}},
  \bibinfo{journal}{ApJ} \textbf{\bibinfo{volume}{573}}, \bibinfo{pages}{754}
  (\bibinfo{year}{2002}).

\bibitem[{\citenamefont{{R. {Shafee}, J. C. {McKinney}, R. {Narayan}, A.
  {Tchekhovskoy}, C. F. {Gammie}, J. E. {McClintock}}}(2008)}]{shafee2008}
\bibinfo{author}{\bibnamefont{{R. {Shafee}, J. C. {McKinney}, R. {Narayan}, A.
  {Tchekhovskoy}, C. F. {Gammie}, J. E. {McClintock}}}}, \bibinfo{journal}{ApJ}
  \textbf{\bibinfo{volume}{687}}, \bibinfo{pages}{L25} (\bibinfo{year}{2008}).

\bibitem[{\citenamefont{{L. G. {Fishbone} \& V. {Moncrief}
  }}(1976)}]{fishbone1976}
\bibinfo{author}{\bibnamefont{{L. G. {Fishbone} \& V. {Moncrief} }}},
  \bibinfo{journal}{ApJ} \textbf{\bibinfo{volume}{207}}, \bibinfo{pages}{962}
  (\bibinfo{year}{1976}).

\bibitem[{\citenamefont{{M. {Koz{\l}owski}, M. {Jaroszy\'nski}, M. A.
  {Abramowicz}}}(1978)}]{kozlowski1978}
\bibinfo{author}{\bibnamefont{{M. {Koz{\l}owski}, M. {Jaroszy\'nski}, M. A.
  {Abramowicz}}}}, \bibinfo{journal}{A\&A} \textbf{\bibinfo{volume}{63}},
  \bibinfo{pages}{209} (\bibinfo{year}{1978}).

\bibitem[{\citenamefont{{M. {Abramowicz} , M. {Jaroszynski}, M.
  {Sikora}}}(1978)}]{abramowicz1978}
\bibinfo{author}{\bibnamefont{{M. {Abramowicz} , M. {Jaroszynski}, M.
  {Sikora}}}}, \bibinfo{journal}{A\&A} \textbf{\bibinfo{volume}{63}},
  \bibinfo{pages}{221} (\bibinfo{year}{1978}).

\bibitem[{\citenamefont{{N. I. {Shakura} \& R. A.
  {Sunyaev}}}(1973)}]{shakura1973}
\bibinfo{author}{\bibnamefont{{N. I. {Shakura} \& R. A. {Sunyaev}}}},
  \bibinfo{journal}{A\&A} \textbf{\bibinfo{volume}{24}}, \bibinfo{pages}{337}
  (\bibinfo{year}{1973}).

\bibitem[{\citenamefont{{R. {Narayan} \& I. {Yi}}}(1994)}]{narayan1994}
\bibinfo{author}{\bibnamefont{{R. {Narayan} \& I. {Yi}}}},
  \bibinfo{journal}{ApJ} \textbf{\bibinfo{volume}{428}}, \bibinfo{pages}{13}
  (\bibinfo{year}{1994}).

\bibitem[{\citenamefont{{R. {Narayan} \& I. {Yi}}}(1995)}]{narayan1995}
\bibinfo{author}{\bibnamefont{{R. {Narayan} \& I. {Yi}}}},
  \bibinfo{journal}{ApJ} \textbf{\bibinfo{volume}{452}}, \bibinfo{pages}{710}
  (\bibinfo{year}{1995}).

\bibitem[{\citenamefont{{S. {Mineshige} \& K.
  {Watarai}}}(2005)}]{mineshige2005}
\bibinfo{author}{\bibnamefont{{S. {Mineshige} \& K. {Watarai}}}},
  \bibinfo{journal}{ChJAA Suppl.} \textbf{\bibinfo{volume}{5}},
  \bibinfo{pages}{49} (\bibinfo{year}{2005}).

\bibitem[{\citenamefont{{R. {Shafee}, J. E. {McClintock}, R. {Narayan}, S. W.
  {Davis}, L.-X. {Li}, R. A. {Remillard}}}(2006)}]{shafee2006}
\bibinfo{author}{\bibnamefont{{R. {Shafee}, J. E. {McClintock}, R. {Narayan},
  S. W. {Davis}, L.-X. {Li}, R. A. {Remillard}}}}, \bibinfo{journal}{ApJ}
  \textbf{\bibinfo{volume}{636}}, \bibinfo{pages}{113} (\bibinfo{year}{2006}).

\end{thebibliography}

\end{document}